# Links between Entropy, Complexity, and the Technological Singularity

Theodore Modis[*]

[*] Address reprint requests to:
Theodore Modis
Growth Dynamics
Via Selva 8
6900 Massagno
Lugano, Switzerland.
Tel. 41-91-9212054, E-mail: tmodis@yahoo.com

## Abstract

Entropy always increases monotonically in a closed system but complexity increases at first and then decreases as equilibrium is approached. Commonsense information-related definitions for entropy and complexity demonstrate that complexity behaves like the time derivative of entropy, which is proposed here as a new definition for complexity. A 20-year old study had attempted to quantify complexity (in arbitrary units) for the entire Universe in terms of 28 milestones, breaks in historical perspective, and had concluded that complexity will soon begin decreasing. That conclusion is now corroborated by other researchers. In addition, the exponential runaway technology trend advocated by supporters of the singularity hypothesis—which was in part based on the trend of the very 28 milestones mentioned above—would have anticipated five new such milestones by now, but none have been observed. The conclusions of the 20-year old study remain valid: we are at the maximum of complexity and we should expect the next two milestones at around 2033 and 2078.

## 1. Introduction

This work was triggered by the author's invitation to speak at the international symposium on *Social singularity in the 21st century: At the crossroads of history* in Prague, CZ on September 18, 2021 (InstituteH21, 2021.) They asked him for an update of his 20-year old work on the evolution of complexity and change in our lives (Modis, 2002; Modis, 2003) and its impact on the possibility of an approaching technological singularity. The author has previously published three related updates (Modis, 2006; Modis, 2012; Modis, 2020.)

During the last ten years there has been much literature published on the subjects of complexity and singularity. One notable example is the work of theoretical physicist Sean M. Carroll whose bestselling book *The Big Picture: On the Origins of Life, Meaning, and the Universe Itself* argues that complexity is related to entropy and that "complexity is about to begin declining" (Carroll, 2016). The idea that complexity first increases and then decreases as entropy increases in closed systems had been previously suggested by several researchers (Huberman et al., 1986; Grassberger, 1989; Li, 1991; Gell-Mann, 1994; Carroll, 2010; Carroll, 2016). In the same direction Kauffman had coined the term "complexity catastrophe" to explain the low complexity of an overly connected network similar to that of a sparsely connected network (Kauffman, 1995). But in a more recent publication, Carroll together with Aaronson and Ouellette demonstrated quantitatively the phenomenon of decreasing complexity when approaching equilibrium by calculating the complexity and the entropy in a cup of coffee that is undergoing the mixing of coffee and cream (Aaronson et al., 2014). These publications provided fertile ground for the work presented here. Two short videos by Sean Carroll popularize these ideas in YouTube for the layperson (Carroll, 2021).

Entropy and complexity are subjects that have enjoyed enormous attention in the scientific literature. Their treatment in the next section is very brief and relates only to their connection to the concept of a technological singularity. With information-related definitions for entropy and complexity, a simple mathematical relationship between them is established in light of which the author reinstates his 20-year old conclusion, namely that we should expect a decreasing complexity in the future instead of an approaching technological singularity. This conclusion has been corroborated by Magee and Devezas who studied shorter-timescale technologically-driven or simply human-driven profound societal changes (Magee et al., 2011).

## 2. Entropy and Complexity

### *2.1 Entropy*

There are many definitions of entropy. The concept was first developed by Rudolf Clausius, a German physicist in the mid-nineteenth century (Clausius, 1867). The classical thermodynamic entropy is defined in terms of the energy (heat) and the temperature of a system. Boltzmann's definition involves the number of different ways the atoms or molecules of a thermodynamic system can be arranged; his celebrated formula for entropy has been carved on his gravestone (Allen et al., 2017). The definition of Gibbs involves the energy and the probability that it occurs for all microstates of the system (Klein, 1990). There is also the quantum-mechanical entropy defined by von Neumann (Zyczkowski et al., 2006). All these definitions of entropy are related to each other but they are not relevant here.

In this paper we will concentrate on the fact that entropy is "a measure of the number of different ways a set of objects can be arranged" or "a measure of disorder" (Martin et al., 2013), even though entropy isn't always disorder (Styer, 2019).[1] With disorder defined as the number of possible configurations, a messy or disordered room has higher entropy than a tidy room. The number of possible configurations of the items in a messy or disordered room is higher than the number of possible configurations in a tidy room, where the items "inhabit a small set of possible places – the books on the bookshelf, the clothes in the dresses, and so on" (Martin et al., 2013).

"The concepts of entropy and disorder are inherently linked" (Martin et al., 2013). When entropy is high disorder is generally high and vice versa. Entropy always increases in a closed system in accordance with the 2nd law of thermodynamics, which stipulates that the entropy S will always increase: $\Delta S > 0$. Entropy may locally decrease, but it will increase elsewhere in the system by at least the same amount so that in a closed system entropy (and also disorder) will generally increase.

There is a link between entropy and information. The higher the number of possible configurations in a system, the more information is needed to describe the system, i.e. the higher its information content will be. In information theory Shannon has defined entropy as a measure of the information content in a message (Shannon, 1948). This is the amount of information an observer could expect to obtain from a given message. A highly ordered, low-entropy state contains less information compared to a highly disordered, high-entropy state. Let's go back to the tidy-room example. If they tell us a living-room is tidy (ordered), the information content of the message is limited. Probably there is a sofa with pillows on it, there is an easy chair, a television against the wall, chairs around a table, etc. But if they tell us that the living room is utterly disordered, the information content of the message is much higher, because it may include oddball situations like pillows on the floor, the television upside down, dirty dishes on the table, chairs scattered around, etc. The more disordered the living room, the greater the information content of the message we are given.

For the rest of this paper we will define entropy as information content.

On a larger scale entropy began increasing at the beginning of the Universe with the Big Bang, when the Universe is thought to have been a smooth, hot, rapidly expanding plasma and rather orderly; a state with low entropy and low information content. Entropy will reach a maximum at the end of the Universe, which in a prevailing view will be a state of heat death, after black holes have evaporated and the acceleration of the Universe has dispersed all energy and particles uniformly everywhere (Carroll, 2010). The information content of this final state of maximal disorder (everything being everywhere), namely the knowledge of the precise position and velocity of every particle in it will also reach a maximum.

Entropy's trajectory grew rapidly during early Universe. As the Universe expansion accelerated, entropy's growth accelerated. Its trajectory followed a rapidly rising exponential-like growth pattern. At the other end, heat death, entropy will grow slowly to asymptotically reach the ceiling of its final maximum (Patel, 2019). It will most likely happen along another exponential-like pattern. It follows that the overall trajectory of entropy will trace some kind of an S-shaped curve with an inflection point somewhere around the middle.

[1] In recent times there has been criticism of the long-standing association of disorder with entropy. The interested reader can go in more depth on this subject by consulting such publications as Floyd, 2007; Lambert, 2002; Low, 1988; Styer 2020; and Wright, 1970.

*2.2 Complexity*

There are also many definitions for complexity. In fact, John Horgan in his essay in his June 1995 *Scientific American* editorial entitled "From complexity to perplexity", has mentioned a list of 31 definitions of complexity (Hogan, 1995). Among them notable is the Kolmogorov complexity, which defines it as a measure of the computational resources needed to specify the object (Kolmogorov, 1963; Kolmogorov 1998). Also, the Effective complexity, defined by Murray Gell-Mann and Seth Lloyd as a measure of the amount of non-random information in a system (Gell-Mann et al., 1996).

But in this paper, and for the sake of consistency with the previous section, we will use the following information-related definition for complexity: the capacity of a system to incorporate information at a given time. Complexity is more like a snapshot while entropy is more like a sum. Informally, complexity reflects the amount of information needed to describe everything "interesting" about the system at a given point in time ("interesting" information is non-random information.) More intuitively, complexity reflects how easy it is to describe the human system; the higher the complexity, the more difficult it is to describe.[2]

In a closed system, entropy and complexity increase together initially, in other words the greater the disorder the more difficult it is to describe the system. But things change later on. Toward the end, as entropy approaches its final maximum where there is also maximal disorder, complexity diminishes. Maximal disorder is simple to describe. By the time entropy reaches its final ceiling the information content has become maximal but also not "interesting" because it has become 100% random information. The degradation of the information content into non-interesting random information begins when entropy reaches the inflection point of its trajectory, i.e. when the rate of growth becomes maximal. At that point complexity goes over a maximum and begins decreasing. Aaronson et al. have likened complexity to "interestingness." They have demonstrated that it declines as entropy reaches a ceiling with the example of a cup of coffee with cream (Aaronson et al., 2014). In the beginning when the cream rests calmly on top of the coffee, the entropy of the system is small (there is also order) and the complexity is also small because the situation is very easy to describe. At the end of the stirring when coffee and cream are completely mixed together, entropy is maximal (there is also maximum disorder because everything is everywhere) but the situation is again easy to describe, so the complexity is low again. Around the middle of the mixing process when entropy (and also disorder) is growing fastest the complexity of the system is maximal.

Another example is the Universe itself. The very early Universe near the Big Bang was a low-entropy and easy to describe state (low complexity.) But the high-entropy state of the end will also be easy to describe because everything will be uniformly distributed everywhere. Complexity was low at the beginning of the Universe and will be low again at the end. It becomes maximal—most difficult to describe—around the middle, the inflection point of entropy's trajectory, when entropy's rate of change is maximal (see milestone numbers 27, 28 in next section.) Complexity follows a bell-shaped curve similar to the time derivative of a logistic function.

[2] This echoes Rosen's epistemological account of complexity: "To say that a system is complex … is to say that we can describe the same system in a variety of distinct ways …" (Rosen, 2000).

*2.3 A new relationship between entropy and complexity*

With the above-mentioned information-related definitions for entropy and complexity for a closed system, namely:

Entropy: the information content
(or a measure of the amount of disorder)
Complexity: the capacity to incorporate information at a given time
(or a measure of how difficult it is to describe at a given time)

we see that entropy results from the accumulation of complexity, or alternatively, that complexity is the time derivative of entropy. Entropy traces out an S-shaped curve while complexity traces a bell-shaped curve. The "interestingness" of entropy's information content diminishes during the second half of the growth process and so does the complexity of the system. At the end there is purely random information everywhere and zero capacity to incorporate "interesting" information.

In this case—i.e. with the chosen definitions—a new relationship between entropy and complexity can be written as:

$$C = \frac{dS}{dt} \qquad (1)$$

or

$$S = \int C \cdot dt \qquad (2)$$

The patterns of the trajectories followed by entropy and complexity may turn out not to be exactly the classical logistic patterns, which are symmetric around the midpoint. But in the coffee-and-cream study mentioned earlier, and with the particular quantitative definitions the investigators used, they found indeed complexity to trace a symmetric bell-shaped curve while entropy approached a ceiling asymptotically, see Figure 2 in (Aaronson et al., 2014).

## 3. Forecasting Complexity

In his 2002 article the author attempted to quantify the evolution of complexity in the Universe in terms of 28 "canonical" milestones—events of maximum importance, breaks in historical perspective—based on data he collected from thirteen different sources (Modis 2002; Modis 2003). In his book *The Singularity Is Near* Kurzweil presented the data behind these 28 milestones in different ways demonstrating the rapid rate of change in our lives, see four figures on pp 17-20 of his book. Together with other runaway trends Kurzweil arrived at the conclusion that there is an approaching technological singularity (Kurzweil, 2005).

These 28 "canonical" milestones generally consist of clusters of events. They are reproduced here in Appendix A. The importance of each milestone was assumed to be proportional to the amount of complexity it brought multiplied by the length of the following stasis until the next milestone. Consequently the increase in complexity $\Delta C_i$ associated with milestone $i$ of importance $I$ is:

$$\Delta C_i = \frac{I}{\Delta T_i} \qquad (3)$$

where $\Delta T_i$ is the time period between milestone *i* and milestone *i*+1.
Under the assumption that milestones of *maximum* importance were also milestones of *comparable* (see *equal*) importance, values for complexity were obtained for 27 milestones in relative terms (i.e. with arbitrary units) as being inversely proportional to the time difference from one milestone to the next one.

In view of the discussion in Section 2.3 the accumulation of this complexity—i.e. the integral—should be akin (if not equal) to the system's entropy. The evolution of the world seen by these 28 milestones is a non-equilibrium open system and for such systems Grandy has demonstrated that it is the time derivative of entropy rather than entropy itself, which plays the major role governing the ongoing macroscopic processes (Grandy, 2004).

Below are reproduced some results from the author's work of twenty years ago. Figure 1 shows the "primordial" S-curve, a logistic fit (thick gray line) to the cumulative complexity values, which should be akin (if not equal) to the entropy of the system. Figure 2 shows complexity per milestone and the fitted curve here (thick gray line) is the bell-shaped logistic life cycle, i.e. the derivative of the logistic function.

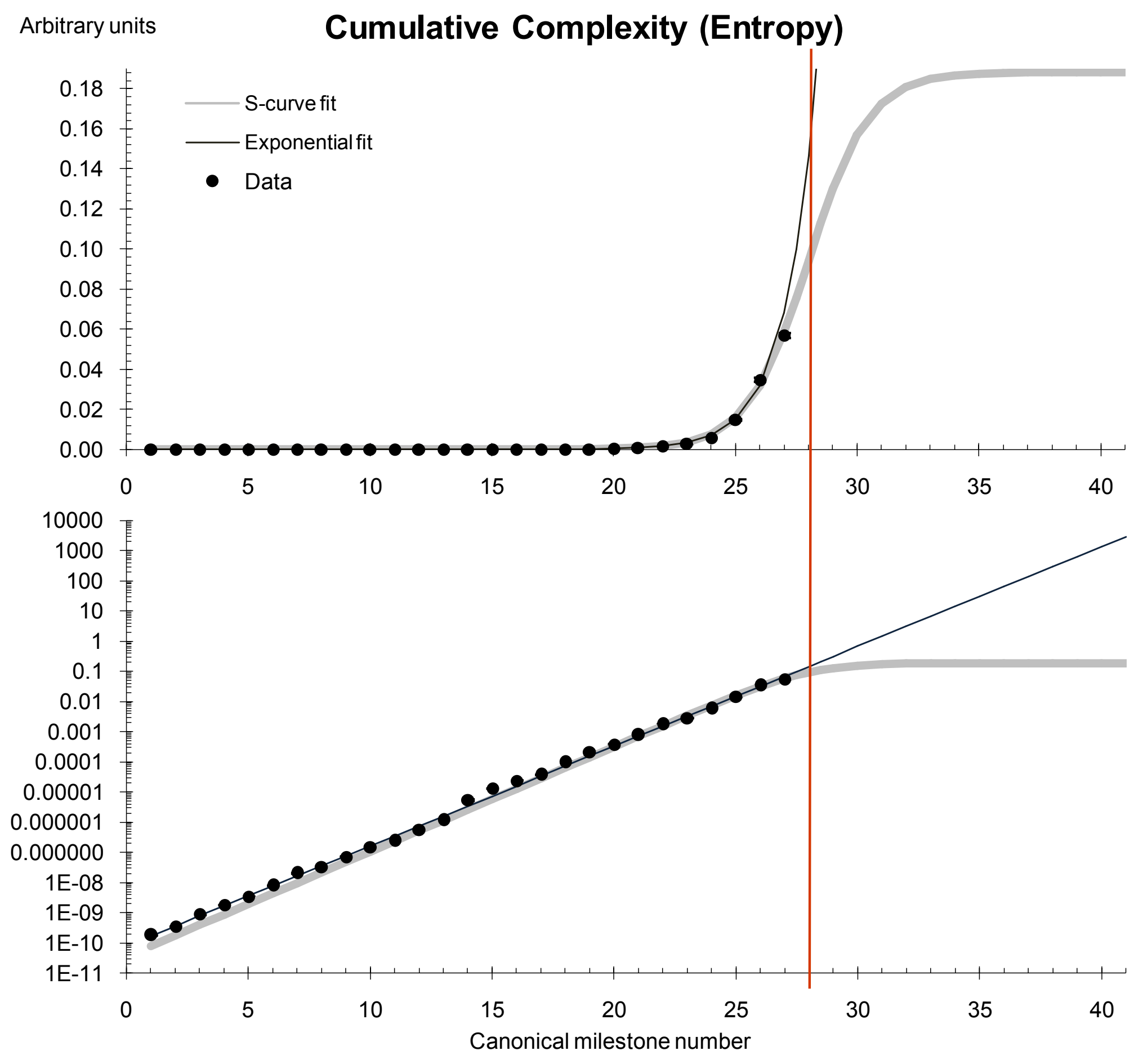

Figure 1. A logistic fit (thick gray line) and an exponential fit (thin black line) to the cumulated complexity values of 27 milestones.The graph at the bottom has a logarithmic vertical scale. The red line is on the 28th milestone and coincides with the center of the logistic.

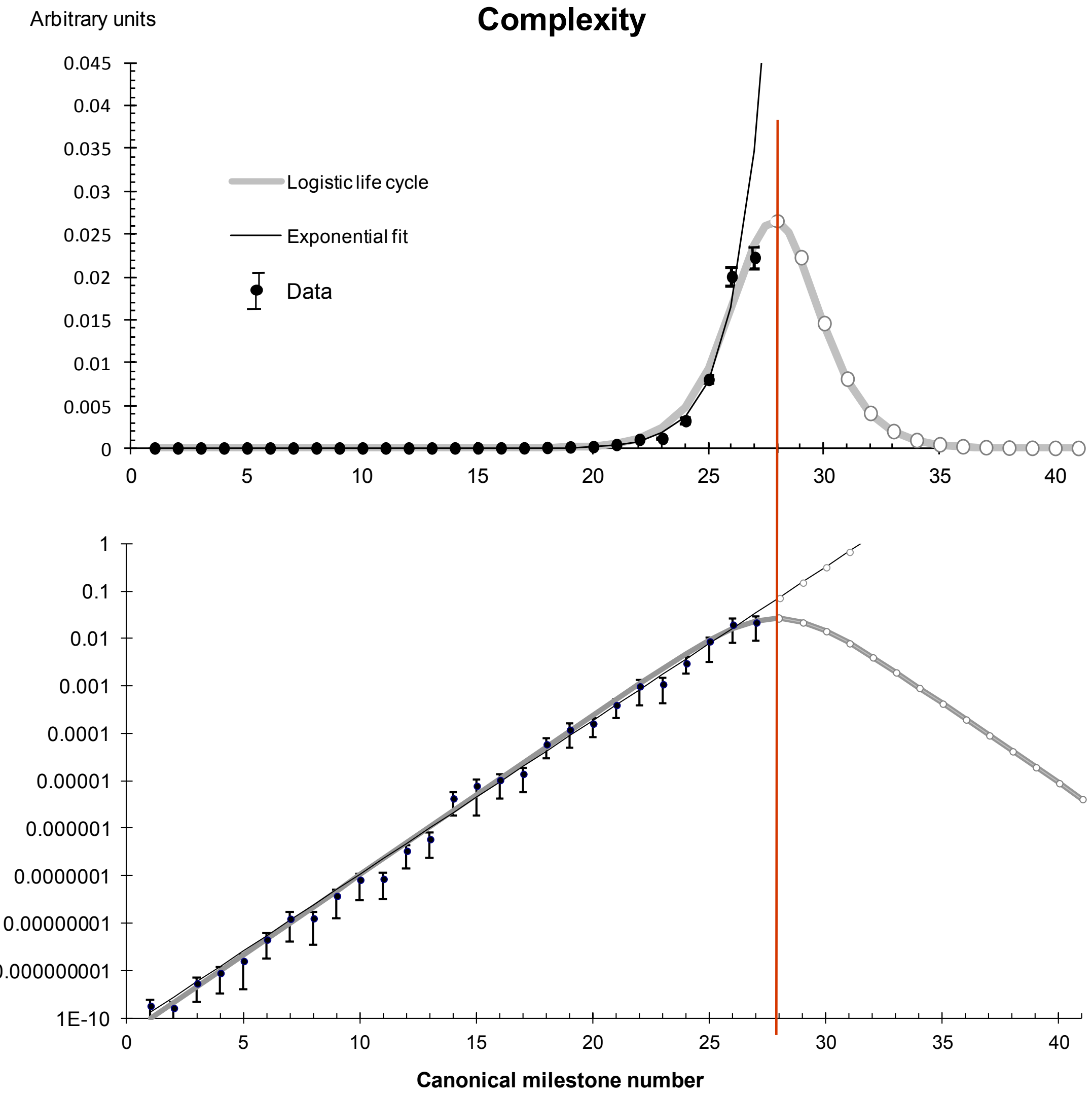

Figure 2. A logistic life-cycle fit (thick gray line) and an exponential fit (thin black line) to the complexity values of 27 milestones. The error bars reflect the spread on the values of the milestones in the particular cluster. The little open circles forecast the position of future milestones according to a logistic and to an exponential extrapolation. The graph at the bottom has a logarithmic vertical scale. The red line is on the 28$^{th}$ milestone and coincides with the center of the logistic.

The red line indicates the 28$^{th}$ milestone for which a complexity value cannot be assigned yet not knowing the 29$^{th}$ milestone. The penetration level of the fitted logistic curve at this time (1990) is 50.1%.

We also see in these two figures an exponential fit to the data (thin black line), which would be compatible with the hypothesis of an approaching singularity. The two fits seem to describe the data comparably well with exception the most recent data point, which is overestimated by the exponential fit, something more obvious in Figure 2.

The little open circles in Figure 2 forecast complexity values for future milestones according to a logistic and to an exponential extrapolation. Since complexity was calculated as being inversely proportional to the time to the next milestone, the forecasted complexity of future milestones—be it with a logistic or an exponential fit—can be translated to dates using Equation (3). Table 1 gives time estimates for the next five milestones according to the two forecasting methods.

**Table 1. Milestone Forecasts**

| Milestone Number | Logistic fit | | Exponential fit | |
|---|---|---|---|---|
| | Complexity* | Year | Complexity* | Year |
| 29 | 0.0223 | 2033 | 0.1540 | 2009 |
| 30 | 0.0146 | 2078 | 0.3247 | 2015 |
| 31 | 0.0081 | 2146 | 0.6846 | 2018 |
| 32 | 0.0041 | 2270 | 1.4435 | 2020 |
| 33 | 0.0020 | 2515 | 3.0436 | 2021 |

* In arbitrary units

## 4. Discussion

Twenty years after the authors original work, his conclusion that complexity and change in our lives will soon begin decreasing is corroborated. First by the work of other scientists who not only claim that complexity in a closed system must eventually decrease, but have also demonstrated with quantitative calculations that it does so symmetrically (Aaronson et al., 2014; Carroll, 2016). And second by the mere fact that no milestones of paramount importance—breaks in historical perspective—have been observed, while five of them had been expected during these twenty years according to the exponential rate of growth advocated by supporters of the singularity hypothesis.

The relationship between entropy and complexity as expressed by Equations (1) and (2) is a direct consequence of the definitions used in Section 2.3, but its validity could be more general despite the fact that the relationship between entropy and complexity is not always one-to-one, as Wentian Li has demonstrated (Li, 1991). As we said earlier the various definitions of entropy are related to each other and so are most of the definitions of complexity. Seeing complexity as the derivative of entropy may have widespread appeal and utility on an intuitive level. After all, complexity reaches a maximum value when entropy grows the fastest. Grandy has amply demonstrated the importance of the role played by the derivative of entropy (Grandy, 2004).

In any case complexity, as determined by the 28 milestones, has reached a maximum and now begins on the declining slope of its bell-shaped pattern. It is a direct consequence of having described the accumulation of entropy by a natural-growth (logistic) pattern, which so far seems to hold as there haven't been any "milestones" in the last 25 years. There have been many small ones but nothing like the Internet, DNA, or nuclear energy. The idea that our world's complexity will decrease in the future may seem difficult to accept but such a unimodal pattern (namely low at the beginning and at the end but high in between, not unlike the normal—Gaussian—distribution[3]) is commonplace in everyday life. It is associated with a reversal appearing at extremes. We say for example, that too much of a good thing is not good. We saw that too much disorder is easy to describe in the examples of coffee and cream, and in the evolution of the entire Universe. Also, I mentioned how Kauffman points out that an overly connected network is as dysfunctional as a sparsely connected network. John Casti in his book X-Events defines complexity as "the number of independent decisions a decision-maker can make at any given time" (Casti, 2012). Thus, if a decision-maker has only few decisions in his or her set of possibilities, he/she faces low complexity. The complexity will increase as the number of possibilities increases. But I believe—Casti does not say this—that if the decision-maker faces millions of possibilities, life in fact will become simpler rather than more complicated because the situation will trigger alternative ways to make decisions (e.g. random choices). Life may not be as simple as having only one choice, but it will be simpler than having to choose among 20 or 30 possibilities, each of which requires individual attention.

[3] The Gaussian and the derivative of the logistic function, the so-called life cycle are very similar (Modis, 2006).

Because the time frame considered by this analysis is vast and the crowding of milestones in recent times is extremely dense functions such as logistics and exponentials cannot describe the growth process adequately. There are processes for which our Euclidean (linear) conception of time does not accommodate an appropriate description. That's why for this analysis, a better-suited time variable was chosen: the sequential milestone number, which is a logistic time scale.

We are obviously dealing with an "anthropic" Universe here since we are overlooking how complexity has been evolving in other parts of the Universe. Still, the author believes that such an analysis carries more weight than just the elegance and simplicity of its formulation. John Wheeler has argued that the very validity of the laws of physics depends on the existence of consciousness.[4] In a way, the human point of view is all that counts! In astronomy/cosmology this is referred to as the Anthropic Principle (Bostrom, 2010), which in its weak form basically states that one sapient life form (humans) looks back to the past from its point of view (Penrose, 1989).

One may object to including such cosmic events as the Big Bang and the formation of galaxies in the same set of milestones as the invention of agriculture, or the internet. But if we dropped the first two milestones and repeated our analysis beginning with the 3rd milestone cluster (the formation of our solar system and the earth, oldest rocks, and origin of life on earth), then the fitted curves would change only imperceptibly. But at the same time, there would now be rough corroboration of the conclusion that complexity and entropy are presently around their midpoints: the sun is close to its midlife (is thought to be 4.6 billion years old and expected to go out in 5.5 billion years from now.)

But we could restrict further our data set to those milestones that have to do only with humans. The reader's attention is drawn to the fact that the trends in Figures 1 and 2 remain purely exponential (straight line on the lower graphs with the logarithmic vertical scales) with extremely low values for most of the range. The trends begin deviating from exponential only very recently, namely after milestone No. 23, i.e. after the fall of Rome, and zero and decimals invented. So even if we dropped all pre-human milestones, we wouldn't obtain a significantly different fit.

One of the thirteen data sets used to distill the 28 "canonical" milestones of Figures 1 and 2 has been provided by Nobel Laureate, Paul D. Boyer. In his contribution he had anticipated two future milestones without specifying their timing. Boyer's 1st future milestone was "Human activities devastate species and the environment," and the 2nd was "Humans disappear; geological forces and evolution continue." The logistic-fit time estimates for the two next milestones from Table 1 are 2033 and 2078 respectively. It is likely that there are *bona fide* scientists who would agree more with Boyer's future milestones and these time estimates rather than with an approaching technological singularity.

Alternatively, and on a more positive and realistic tone the next two milestones could well be along the lines:

---

[4] John Wheeler was a renowned American theoretical physicist best known for first using the term "black hole" in 1967.

- 2033. A cluster of achievements in AI, robotics, nanotechnology, bioengineering, NASA's scheduled human mission to Mars, etc. could qualify as one milestone in the same way modern physics, radio, electricity, automobile, and airplane had done at the turn of the twentieth century (milestone No. 26).
- 2078. Teleportation or creation of life, two fields that have been attracting attention of researchers for some time now.

In his publication of 2002 the author had concluded that "we are sitting on top of the world" from the point of view that we are experiencing complexity and change at their maximum and that they will begin decreasing soon. Twenty years later there is no reason to revise that conclusion.

**Author statement**

I would like to thank Alain Debecker and Athanasios G. Konstandopoulos for fruitful discussions.

## Appendix A

The 28 "canonical" milestones generally represent an average of clustered events not all of which are mentioned in this table. That is why some events, e.g. the asteroid collision, may appear dated somewhat off. Highlighted in bold are in the most outstanding event in the cluster. The dates given are expressed in number of years before year 2000.

| No. | Milestone | Date |
|---|---|---|
| 1. | **Big Bang** and associated processes | $1.55 \times 10^{10}$ |
| 2. | **Origin of Milky Way,** first stars | $1.0 \times 10^{10}$ |
| 3. | **Origin of life on Earth,** formation of the solar system and the Earth, oldest rocks | $4.0 \times 10^{9}$ |
| 4. | **First eukaryotes**, invention of sex (by microorganisms), atmospheric oxygen, oldest photosynthetic plants, plate tectonics established | $2.1 \times 10^{9}$ |
| 5. | **First multicellular life** (sponges, seaweeds, protozoans) | $1.0 \times 10^{9}$ |
| 6. | **Cambrian explosion,** invertebrates, vertebrates, plants colonize land, first trees, reptiles, insects, amphibians | $4.3 \times 10^{8}$ |
| 7. | **First mammals,** first birds, first dinosaurs, first use of tools | $2.1 \times 10^{8}$ |
| 8. | **First flowering plants,** oldest angiosperm fossil | $1.3 \times 10^{8}$ |
| 9. | **Asteroid collision,** first primates, mass extinction, (including dinosaurs) | $5.5 \times 10^{7}$ |
| 10. | **First hominids,** first humanoids | $2.85 \times 10^{7}$ |
| 11. | **First orangutans,** origin of proconsul | $1.66 \times 10^{7}$ |
| 12. | **Chimpanzees and humans diverge,** earliest hominid bipedalism | $5.1 \times 10^{6}$ |
| 13. | **First stone tools,** first humans, Ice Age, *Homo erectus*, origin of spoken language | $2.2 \times 10^{6}$ |
| 14. | **Emergence of *Homo sapiens*** | $5.55 \times 10^{5}$ |
| 15. | **Domestication of fire,** *Homo heidelbergensis* | $3.25 \times 10^{5}$ |
| 16. | **Differentiation of human DNA types** | $2.0 \times 10^{5}$ |
| 17. | **Emergence of "modern humans,"** earliest burial of the dead | $1.06 \times 10^{5}$ |
| 18. | **Rock art, protowriting** | $3.58 \times 10^{4}$ |
| 19. | **Invention of agriculture** | $1.92 \times 10^{4}$ |
| 20. | **Techniques for starting fire,** first cities | $1.1 \times 10^{4}$ |
| 21. | **Development of the wheel, writing** | 4907 |
| 22. | **Democracy,** city-states, the Greeks, Buddha | 2437 |
| 23. | **Zero and decimals invented,** Rome falls, Moslem conquest | 1440 |
| 24. | **Renaissance (printing presss),** discovery of New World, the scientific method | 539 |
| 25. | **Industrial revolution (steam engine),** political revolutions (France, USA) | 223 |
| 26. | **Modern physics,** radio, electricity, automobile, airplane | 100 |
| 27. | **DNA structure described, transistor invented, nuclear energy**, World War II, Cold War, Sputnik | 50 |
| 28. | **Internet, human genome sequenced** | 5 |

**Biography**

Theodore Modis is a physicist, strategist, futurist, and international consultant. He is author/co-author to over one hundred articles in scientific and business journals and ten books. He has on occasion taught at Columbia University, the University of Geneva, at business schools INSEAD and IMD, and at the leadership school DUXX, in Monterrey, Mexico. He is the founder of Growth Dynamics, an organization specializing in strategic forecasting and management consulting: http://www.growth-dynamics.com